\documentclass[a4paper]{jpconf}
\usepackage{graphicx}
\begin{document}
\title{Two Higgs doublet model and leptoquarks constraints from D meson decays}

\author{J. Barranco, D. Delepine, V. Gonzalez Macias}
\address{Departamento de F\'isica, Divisi\'on de Ciencias e Ingenier\'ia, Campus Le\'on, Universidad de Guanajuato}
\author{L. Lopez-Lozano}
\address{\'Area Acad\'emica de Matem\'aticas y F\'{\i}sica,
  Universidad Aut\'onoma del Estado de Hidalgo,
  Carr. Pachuca-Tulancingo Km. 4.5, C.P. 42184, Pachuca, Hgo.}

\begin{abstract}
We use a combined analysis of the semileptonic and leptonic branching ratios of the D mesons to constrain scalar
leptoquark interactions and charged higgs-like interactions.
For the THDM type II, we found that a low mass $6.3\mathrm{GeV} <m_{H^{+}}<63.1\mathrm{GeV}$ for the charged Higgs is favored at 90\% C.L.
although at 95\% there is still agreement with other constraints. We find for the leptoquark states a more restrictive bound than
previous analysis.
\end{abstract}
\section{Introduction}
The intensity frontier \cite{Hewett:2012ns,Cirigliano:2013lpa} is a low
energy probe and indirect search of New Physics (NP). One could search for forbidden or highly suppressed processes
in the SM, such as electric dipole moments, lepton number violation,
flavor changing neutral processes, among others. Another possibility is to analyze measured quantities
allowed within the SM but with sufficient efficiency to test non standard interactions (NSI).

The physics of the D meson has gained recent importance in intensity frontier mainly due to improvement
in the theoretical calculation, coming from lattice, of the form factors for $D\to K$ in semileptonic decays
\cite{Koponen:2013tua}. Likewise, the experimental measurements that shapes the model-independent form factor with
small uncertainty given by CLEO \cite{Ge:2008aa} demonstrates a good agreement with lattice.
Thus, slight deviations seen in the SM branching ratios (BR) for semileptonic D decays could be an indirect signal
of physics beyond the standard model (BSM)
coming mainly from off-shell non standard particles, rather than errors in calculation of the form factors.

Due to the lack of a fundamental theory, an effective approach is useful because it allows us to found bounds that can
be later be translated into constraints on the relevant parameters for a huge variety of models.
Hence, the intensity frontier is greatly benefited by the use of effective theories.
This approach has been used by using data mostly from B and K mesons.

On the other hand, only a couple of analysis has been done using the Semileptonic and leptonic decays of the D mesons
\cite{Dobrescu:2008er,Kronfeld:2008gu,Barranco:2013tba}.
In particular, in \cite{Dobrescu:2008er}, the lattice calculations of $f_{Ds}$ implied a $3.8\sigma$ deviation of the SM prediction over the experimental value. Thus provided an strong indication for new physics.
The discrepancy was alleviated once the form factors were re-evaluated and now it is known with a high level of precision (around $2\%$ error) \cite{Davies:2010ip}.

In \cite{Kronfeld:2008gu} it was suggested the possibility of doing a combined analysis of the semileptonic and leptonic decays of the D mesons but the analysis was not performed. Such analysis was recently done \cite{Barranco:2013tba} as suggested in  \cite{Kronfeld:2008gu} now including the updated values of both the semileptonic form factors \cite{Koponen:2013tua} and leptonic decay constant $f_{Ds}$ \cite{Davies:2010ip}. Because this combination of different BRs it is possible to extract bounds for each relevant Wilson coefficient independently. Given the current knowledge of the form factors, we found that the possible evidence of new physics found in
\cite{Dobrescu:2008er} has now gone and instead, it is now possible to extract bounds on models beyond the SM.

One possible source of error in the analysis in \cite{Barranco:2013tba} is the use of the current world average value of $V_{cs}$ \cite{Beringer:1900zz}. In order to relax that assumption and observe if the bounds reported are accurate, we performed here a new analysis where we have used ratios of BRs instead of only the BRs for the $\chi^2$ analysis. Since we have only processes that involve the $c-s$ transition, by using the ratios it is possible to eliminate the functional dependence on $V_{cs}$ and consequently the error involved in the CKM matrix element.
This new analysis implies that: {\it i)} the goodness of the fit given by $\chi^2_{min}/\mbox{d.o.f}$ is worst when we use
the ratios instead of the decay rates, and {\it ii)} given the arbitrary of selecting different ratios, the bounds change. Nevertheless, we will show that under some
selected ratios, there is consistency between the constraints found in \cite{Barranco:2013tba} and those obtained here by means of the ratios.

\section{Branching ratios: Theory and experiment}\label{Section2}

\begin{figure}
\begin{center}
\includegraphics[width=0.9\textwidth]{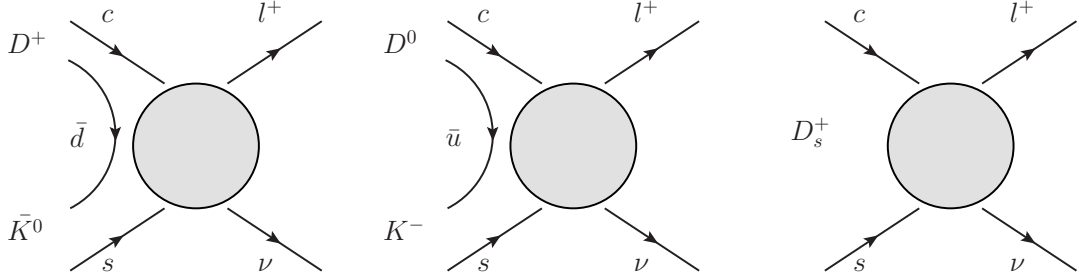}
\caption{Schematic representation of the D meson decays we are interested. Circle denotes any type on non standard interaction described by the
Lagrangian \ref{NSILagrangian}}\label{NSIdiagrams}
\end{center}
\end{figure}

The approach we will use in this work will be that of effective Lagrangians.  
The effective Lagrangian with NSI and standard interactions that describes the $c\to s$ transition is given by
$\mathcal{L}=\mathcal{L}_{SM}+\mathcal{L}_{NP}$
where
\begin{equation}
\mathcal{L}_{SM}=2\sqrt{2}G_FV^{*}_{cs}(\bar{s_{L}}\gamma_{\mu}c_{L})(\bar{\nu}_{L}\gamma^{\mu}l_{L}\,,
\end{equation}
and the Lagrangian that includes all NSI is given by
\begin{eqnarray}
-\frac{\mathcal{L}_{NP}}{G_F}&=&C^{V,LL}_{sc\ell\nu}(\bar{s_{L}}\gamma_{\mu}c_{L})(\bar{\nu}_{L}\gamma^{\mu}l_{L})+C^{V,RL}_{sc\ell\nu}(\bar{s_{R}}\gamma_{\mu}c_{R})(\bar{\nu}_{L}\gamma^{\mu}l_{L})\nonumber \\
&+&C^{S,RR}_{sc\ell\nu}(\bar{s_{L}}c_{R})(\bar{\nu_{L}}l_{R})+C^{S,LR}_{sc\ell\nu}(\bar{s_{R}}c_{L})(\bar{\nu}_{L}l_{R})\nonumber \\
&+&C^{T,LR}_{sc\ell\nu}(\bar{s_{R}}\sigma_{\mu\nu}c_{L})(\bar{\nu}_{L}\sigma^{\mu\nu}l_{R})+C^{T,RR}_{sc\ell\nu}(\bar{s}_{L}\sigma_{\mu\nu}c_{R})(\bar{\nu}_{L}\sigma^{\mu\nu}l_{R})\label{NSILagrangian}
\end{eqnarray}

where the $C$'s are the Wilson coefficients, that parametrize all new type of interactions beyond the Standard Model at a given scale: Vectorial (V), Scalar (S) or tensorial (T) interactions. Notation is the same as \cite{Barranco:2013tba}. In Fig. \ref{NSIdiagrams} are shown the D meson decays we are interested in and in.

Next we will compute the decay rates of the leptonic  $D_s\to \ell \nu_\ell$ decays and the semileptonic decays $D\to K\ell\nu_\ell$ including NSI Lagrangian \ref{NSILagrangian}.

\subsection{$D^{\pm}_s \to l^{\pm} \nu$}
\begin{table}[t]
\begin{center}
\begin{tabular}{l|c|c}
\hline\hline
$i$ Decay& Theo. BR $\mathcal{B}_i^{th}$ & Exp. BR $\mathcal{B}_i^{exp}$ \\
\hline\hline
1 $D^0\to K^- e^+ \nu_e$            & $(3.28\pm0.11)\%$.          & $(3.55\pm0.04)\%$\\
2 $D^0\to K^- \mu^+ \nu_\mu$        & $(3.22\pm0.11)\%$           & $(3.30\pm0.13)\%$\\
3 $D^+\to \bar K^0 e^+ \nu_e$       & $(8.40\pm0.32)\%$.          & $(8.83\pm0.22)\%$\\
4 $D^+\to \bar K^0 \mu^+ \nu_\mu$   & $(8.24\pm0.31)\%$           & $(9.2\pm0.6)\%$\\
5 $D_s^+\to \tau^+ \nu_\tau$         & $(5.10\pm0.22)\%$          & $(5.43\pm 0.31)\%$\\
6 $D_s^+\to \mu^+ \nu_\mu$          & $(5.20\pm0.20)\times10^{-3}$ & $(5.90\pm 0.33)\times10^{-3}$\\
\hline\hline
\end{tabular}
\caption{Theoretical and experimental branching ratios  \cite{Barranco:2013tba}}\label{table1}
\end{center}
\end{table}
Let us consider first the full-leptonic decay of the $D$ meson, $D^{\pm}(p)\to\nu(p_1)l^{\pm}(p_2)$, with effective non-standard interactions. The only non-vanishing hadronic elements for the pseudo-scalar meson are
 \begin{eqnarray}
 \langle 0|\bar{s}\gamma_{\mu}\gamma_{5} c|D(p)\rangle&=&if_{D_s}p_{\mu}\,, \\
 \langle 0|\bar{s}\gamma_{5} c|D(p)\rangle&=&i f_{D_s}\frac{M_{D_s}}{m_{c}+m_{s}}\,.
 \end{eqnarray}

The decay rate of $D_s\to\ell\nu_\ell$ including the SM Lagrangian plus the NSI Lagrangian in the rest frame of the decaying meson is given by

\begin{equation}
\Gamma_{D_s\to\ell\nu}=\frac{|G_{F}f_{D_s}\left(M^{2}_{D_s}-m^{2}_{l}\right)|^2}{8\pi M_{D_s}^3}\left|V_{cs}m_l
+\frac{m_{l}(C^{V,LL}_{sc\ell\nu}-C^{V,RL}_{sc\ell\nu})}{2\sqrt{2}}+\frac{M^{2}_{D_s}(C^{S,RR}_{sc\ell\nu}-C^{S,LR}_{sc\ell\nu})}{2\sqrt{2}(m_c+m_s)}
\right|^2\,.\label{gamma_leptonic}
\end{equation}

On the other hand, in the rest frame of the decaying meson, the partial decay rate for the $D^{0}\to K^{\pm} l^{\mp} \nu$ decay channel with non standard interactions is given by

\textbf{\begin{eqnarray}
&&\frac{d\Gamma_{D\to K\ell\nu_\ell}}{dE_K}=
\frac{G_F^2 m_D \sqrt{E_K^2-m_K^2}}{(2\pi)^3}
\left\{ (E_K^2-m_K^2) \frac{2q^2 + m_\ell^2}{3q^2}
\left|(V^{*}_{cs}+G_V)f_+(q^2)\right|^2 \right.
\nonumber
\\ &+& \left.\left(-|G_Tf_2(q^2)|^2\frac{q^2+2m_l^2}{3}+
m_l(V^{*}_{cs}+G_V)f_+(q^2)G_T^*f_2(q^2)\right)\left(\frac{E_k^2-m_K^2}{m_D^2}\right)
\right. \nonumber\\ &+&
\left.\frac{\left|(m_D^2-m_K^2)qf_0(q^2)\right|^2}{4m_D^2}
\left|\frac{m_\ell}{q^2}(V^{*}_{cs}+G_V)+\frac{G_S}{m_c-m_s}\right|^2
\right\} \left(1-\frac{m_\ell^2}{q^2} \right)^2\,,\label{gamma_semileptonic}
\end{eqnarray}}

where we have defined
$G_V=(C^{V,LL}_{sc\ell\nu}+C^{V,RL}_{sc\ell\nu})/2\sqrt{2}$,
$G_S=(C^{S,RR}_{sc\ell\nu}+C^{S,LR}_{sc\ell\nu})/2\sqrt{2}$ and $G_T=(C^{T,RR}_{sc\ell\nu}+C^{T,LR}_{sc\ell\nu})/2\sqrt{2}$.

\begin{figure}
\begin{center}
\includegraphics[width=0.75\textwidth]{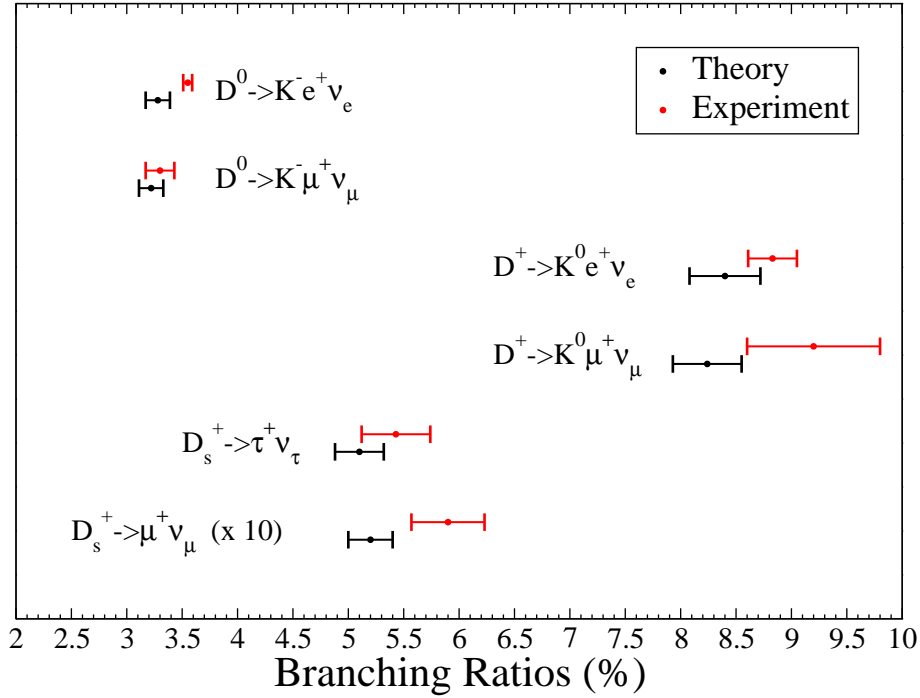}
\caption{Theoretical branching ratios (black lines) and experimental branching ratios of the D meson decays (red lines). }\label{Brtheo}
\end{center}
\end{figure}

The scalar $f_0(q^2)$, vector $f_+(q^2)$ and tensorial $f_2(q^2)$ form factors are defined via the non-vanishing hadronic elements:
\begin{eqnarray}
 < K(k)|\bar{s}\gamma_{\mu}c|D(p)>&=&(p^{\alpha}+k^{\alpha}-\frac{m^{2}_{D}-m^{2}_{K}}{q^{2}}q^{\alpha})f_{+}(q^{2})\nonumber \\&+& \frac{m^{2}_{D}-m^{2}_{K}}{q^{2}}q^{\alpha}f_{0}(q^{2})\,, \\
 < K(k)|\bar{s}\sigma^{\alpha\beta}c|D(p)>&=&im_{D}^{-1} f_{2}(q^2)(p^\alpha k^\beta -p^\beta k^\alpha) \,,\\
 < K(k)|\bar{s}c|D(p)>&=&\frac{m^2_D-m^2_K}{m_c-m_s}f_{0}(q^2)\,.
 \end{eqnarray}
Let us first compute the current theoretical branching ratios. We fix all Wilson coefficients to zero and use the PDG \cite{Beringer:1900zz} average values for Fermi Constant $G_F$, the masses of the quarks $m_c,m_s$ and leptons $m_e,m_\tau,m_\mu$, the CKM element $V_{cs}$ and the masses of the $K$ and $D$ mesons. Furthermore, we use the latest reported semileptonic form factors \cite{Koponen:2013tua} and the leptonic $f_{Ds}$ constants \cite{Davies:2010ip}. Theoretical error bars are computed by propagating the errors in all variables.
The same procedure was done in \cite{Dobrescu:2008er} where they used  the current values for the constants as reported in PDG at that time.
The new current branching ratios are illustrated in Fig. \ref{Brtheo} (black points) for
$D_s\to \ell \nu_\ell$, $D^0\to K^-\ell^+\nu_\ell$ and $D^+\to K^0\ell^+\nu_\ell$.
In the same figure, red points with error bars are the experimental branching ratios.
It can be noted that except for the $D^0\to K^- e^+\nu_e$ and $D_s^+\to \mu^+\nu_\tau$, the rest of the theoretical  branching ratios are in agreement with the experimental ones at $1 \sigma$.
\subsection{Effect of NSI in $D$ meson decays rates.}
\begin{figure}
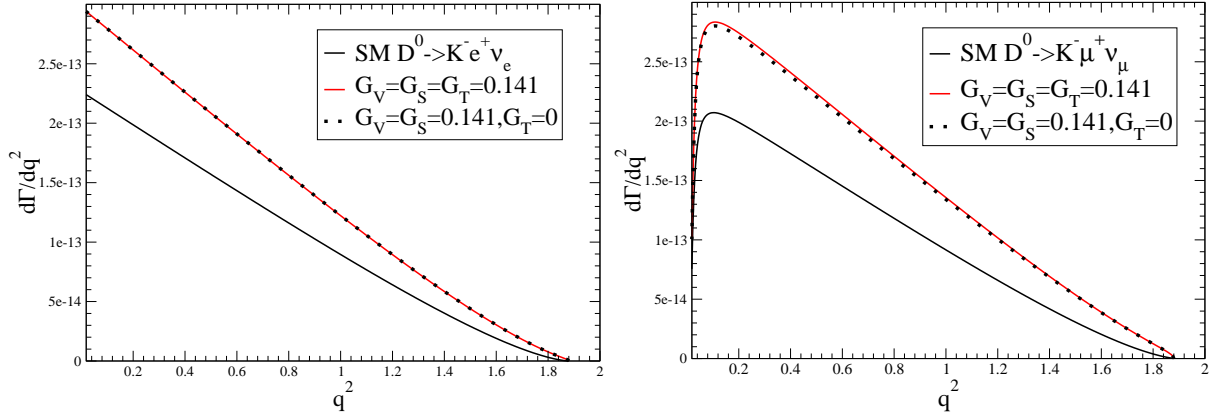

\begin{center}
\includegraphics[width=0.49\textwidth]{tensorial_effect.eps}
\includegraphics[width=0.49\textwidth]{tensorial_effect_muon.eps}
\caption{Theoretical differential decay for $D^0\to K^- e^+\nu_e$ and $D^0\to K^- \mu^+\nu_\mu$ decays. Black solid lines correspond to fixing Wilson Coefficients to zero. Red are decay rates when all NSI interactions are included for the case $C_{cs\ell\nu}^{S,RR}=C_{cs\ell\nu}^{S,LR}=C_{cs\ell\nu}^{V,LL}=
C_{cs\ell\nu}^{V,RL}=C_{cs\ell\nu}^{T,RR}=C_{cs\ell\nu}^{T,LR}=0.2$. Dotted line only includes scalar and vectorial NSI  with $C_{cs\ell\nu}^{S,RR}=C_{cs\ell\nu}^{S,LR}=C_{cs\ell\nu}^{V,LL}=
C_{cs\ell\nu}^{V,RL}=0.2$ and the tensorial NSI are switched off, i.e.$C_{cs\ell\nu}^{T,RR}=C_{cs\ell\nu}^{T,LR}=0$.}\label{dgamma}
\end{center}
\end{figure}

The inclusion of the effective Lagrangian \ref{NSILagrangian} changes the
theoretical estimation of the BRs as can be seen in eqs. \ref{gamma_leptonic}-\ref{gamma_semileptonic}.
In order to illustrate the effect of NSI in the decay rate, we will compute the $D^0\to K^- e^+\nu_e$ and $D^0\to K^- \mu^+\nu_\mu$ decay rate including NSI.
In addition to the Wilson coefficients we need to assume some functional dependence of the tensorail form factor $f_2(q^2)$. For definitiveness we will use a one pole form factor given by
\begin{equation}
f_2(q^2)=\frac{f_2(0)}{\left(1-q^2/m_{D*}^2\right)}\,.
\end{equation}
Currently there is no lattice calculation for $f_2(0)$. Let us for a moment suppose it is $f_2(0)=1$. For illustration purposes let us fix the Wilson coefficients $C_{cs\ell\nu}^{S,RR}=C_{cs\ell\nu}^{S,LR}=C_{cs\ell\nu}^{V,LL}=C_{cs\ell\nu}^{V,RL}
=C_{cs\ell\nu}^{T,RR}=C_{cs\ell\nu}^{T,LR}=0.2$. The differential decay rate $d\Gamma_{D\to K\ell\nu_\ell}/dE_K$  increases can be observed in both graphics of Figure \ref{dgamma}. Black solid line is the differential decay rate when all Wilson coefficients are fixed to zero, i.e. those are the SM theoretical branching ratios. Red solid line are the new decay rate once the Wilson coefficients are included. Let us switch off the tensorial NSI, that is, $C_{cs\ell\nu}^{T,RR}=C_{cs\ell\nu}^{T,LR}=0$ and we let both scalar and vectorial NSI fixed to $C_{cs\ell\nu}^{S,RR}=C_{cs\ell\nu}^{S,LR}=C_{cs\ell\nu}^{V,LL}=C_{cs\ell\nu}^{V,RL}
=0.2$ as before. The result are the dotted points shown in Fig. \ref{dgamma}. One may conclude that the tensorial NSIs have not a strong effect on the decay rates.
Hence, even there is not a theoretical estimation of the form factor, the total effect
of the tensorial contribution when both scalar and vectorial NSI are included is negligible.

\section{Wilson coefficients for THDM and Leptoquarks}
Here we will derive the corresponding Wilson coefficients for two models beyond the SM, namely the THDM and Leptoquark model. The low energy limit of each model can be directly related with a four Fermi point interaction that can be parametrized via a Wilson coefficients. Next we perform such limit for those models.
\subsection{Two Higgs Doublet model}
The interaction between quarks and leptons with scalars in the THDM is described in general by the Yukawa Lagrangian
\begin{equation}\label{2HDm_lagrangian}
-\mathcal{L}_\textrm{Y}^\textrm{THDM}=\bar{Q'}_L\sum_{a=1}^2\left(Y^d_a\Phi_a d_R+Y^u_a\tilde{\Phi}_a u_R\right)+\sum_{a=1}^2\bar{L}_LY^\ell_a\Phi_a\ell_R+\textrm{H.c.}\,,
\end{equation}
where $Q'_L=(u_L,d_L)^T$ and $L=(\nu_L, \ell_L)^T$ are the left-handed $SU(2)$ doublets of quark and leptons respectively. $u_R$, $d_R$ and $\ell_R$ are the right-handed singlets. The Higgs doublets are given by $\Phi_a=(\varphi^+_a,\varphi^0_a)^T$ and $\tilde{\Phi}$ is the charge conjugate doublet. These doublets have a non-zero VEV given by $\langle \Phi_a\rangle=(0,v_a)^T$ where $v^2=v_1^2+v_2^2$. Charged scalar interactions in the versions without Flavor Changing Neutral Currents (FCNC) at the tree level can be described introducing discrete symmetries on the doublets that restrict the Yukawa couplings. In this case the Lagrangian, using the Higgs basis for the charged scalars and the mass basis for fermions, is given by \cite{Barger:1989fj}
\begin{equation}
-\mathcal{L}_{H^\pm}=H^+\left[\frac{\sqrt{2}V_{u_id_j}}{v}\bar{u}_i(m_{u_i}XP_L+m_{d_j}YP_R)d_j+\frac{\sqrt{2}m_\ell}{v}Z\bar{\nu}_L\ell_R\right]+\textrm{H.c.}.
\end{equation}
Here the operators $P_{R,L}=(1\pm\gamma^5)/2$ are the chiral projectors. The parameters $X,Y$ and $Z$ dependency on $\beta$ are given in the Table \ref{table1} for version without FCNC.
\begin{table}[t]
\begin{center}
\caption{Dependencies of the free parameters $X,Y$ and $Z$ in terms of $\beta$ for every version of the THDM without FCNC \cite{Branco:2011iw}}\label{table2}
\begin{tabular}{ccccc}
\hline\hline
&Type I&Type II&LS(Lepton Specific)&Flipped\\
\hline\hline
$X$&$\cot\beta$&$\cot\beta$&$\cot\beta$&$\cot\beta$\\
$Y$&$-\cot\beta$&$\tan\beta$&$-\cot\beta$&$\tan\beta$\\
$Z$&$-\cot\beta$&$\tan\beta$&$\tan\beta$&$-\cot\beta$\\
\hline\hline
\end{tabular}
\end{center}
\end{table}
The effective Lagrangian can be written as
\begin{equation}
\mathcal{L}^{\textrm{THDM}}_{\textrm{eff}}=2\sqrt{2}G_FV_{cs}^*\left(\frac{m_\ell m_c}{M_H^2}ZXQ_3+\frac{m_\ell m_s}{M_H^2}ZYQ_4\right),
\end{equation}
thus we can see that the only non zero Wilson coefficients are:
\begin{equation}
C_{cs\ell\nu}^{S,RR}=2\sqrt{2}V_{cs}^*\frac{m_\ell m_c}{M_H^2}ZX\,,\quad C_{cs\ell\nu}^{S,LR}=2\sqrt{2}V_{cs}^*\frac{m_\ell m_s}{M_H^2}ZY\,.\label{WilsonTHDM}
\end{equation}

\subsection{Leptoquarks}
 Leptoquark particles are scalars or vectors bosons expected to exist in various extensions of the standard model that carry both baryon  and lepton number \cite{Pati:1974yy,Buchmuller:1986zs}. They emerge for instance in grand unified theories (GUTs) \cite{Langacker:1980js,Frampton:1991ay,Hewett:1988xc}, technicolor models \cite{Farhi:1980xs,Lane:1991qh}  and SUSY models with R-parity violation. At low energy, Leptoquarks can be  described as an effective four fermion interaction induced by leptoquark exchange.
 Several observables have been used to set bounds on these effective couplings
as is the case of D meson decays \cite{Davidson:1993qk,Dobrescu:2008er,Dorsner:2009cu}.

Scalar leptoquarks $S$ may couple to both left or right handed quark chiralities.  Let us consider the exchange of the following scalar leptoquarks:
\begin{itemize}
\item $S_{0}$ with charge $-1/3$ and $(3,1,-2/3)$ gauge numbers; and
\item the $S_{1/2}$ with charge $2/3$ and $(3,2,7/3)$ gauge numbers.
\end{itemize}
For this particular case, the effective Lagrangian for the $c\to s$ transition is:

\begin{equation}
 L^{LQ}_{Eff}=V^{*}_{cs}\left[\frac{\kappa^{R*}_{i2}\kappa^{L}_{i2}}{m^{2}_{S_{1/2}^{2/3}}}(\overline{\nu^{\
i}_L}c_R\overline{l^ {c}_{iL}}s^{c}_R) + \frac{\kappa'^{R*}_{i2}\kappa'^{L}_{i2\
}}{m^{2}_{S_{0}^{-1/3}}}(\overline{\nu^{i}_L}s^{c}_R\overline{l^ {c}_{iL}}c_R)+\frac{|\kappa'^{L}_{i2}|^{2}}{\
m^{2}_{S_{0}^{-1/3}}}(\overline{\nu^{i}_L}s^{c}_R\overline{l^ {c}_{iR}}c_L)\right]\,.\label{LeptoquarkLagrangian}
 \end{equation}
Now we will re-write the Lagrangian \ref{LeptoquarkLagrangian} in order to have external quark and lepton currents through a Fierz transformations. The result is:

\begin{eqnarray}
\mathcal{L}^{LQ}_{Eff}&=&\frac{1}{2}V^{*}_{cs}
\left[\left(\frac{\kappa^{R*}_{i2}\kappa^{L}_{i2}}{m^2_{S_{1/2}^{2/3}}}+
\frac{\kappa'^{R*}_{i2}\kappa'^{L}_{i2}}{m^{2}_{S_{0}^{-1/3}}}\right)
\left(\overline{\nu_L}^ {i}l_{iR}\overline{s_L}c_R
+ \frac{1}{4}\overline{\nu_L}^{i}\sigma_{\mu\nu}l_{iR}\overline{s_L}\sigma^{\mu\nu}c_R\right)
\right. \nonumber \\
&-&\left. \frac{|\kappa'^{L}_{i2}|}{m^{2}_{S_{0}^{-1/3}}}\left(\overline{\nu}^{i}\gamma^{\mu}P_L 
l_{i} \overline{s}\gamma_{\mu}P_L c\right) \right]\,.
\end{eqnarray}

that lead to tensor, scalar and vector interactions, that we shall take into account in a model dependent analysis.
Explicitly, the only non-vanishing Wilson coefficients are:
\begin{equation}
C^{VLL}_{sc\ell\nu}=\frac{\sqrt{2}V_{cs}}{G_F}\left(\frac{|\kappa'^{L}_{i2}|}{m^{2}_{S_0^{-1/3}}}\right)
\quad \mbox{and} \quad
C^{TRR}_{sc\ell\nu}=\frac{\sqrt{2}V_{cs}}{G_F}\left(\frac{\kappa^{R*}_{i2}\kappa^{L}_{i2}}{m^{2}_{S_{1/2}^{2/\
3}}}+\frac{\kappa'^{R*}_{i2}\kappa'^{L}_{i2}}{m^{2}_{S_0^{-1/3}}}\right)\label{WilsonLepto}
\end{equation}
and observe that $C^{SRR}_{sc\ell\nu}=-4C^{TRR}_{sc\ell\nu}$.

\section{Constraining new physics with D meson decays}
In the last years, a new level of precision has been achieved in measurements of branching fractions for
leptonic and  semileptonic D decays by the Belle, BaBar, and CLEO collaborations
\cite{Besson:2007aa,Besson:2009uv,Ge:2008aa,Widhalm:2007ws,Link:2004dh,Aubert:2007wg,Dobbs:2007aa}. In Table \ref{table1} we have
collected the world average values of the Branching fractions for the D meson decays as reported by PDG
\cite{Beringer:1900zz}. Furthermore, there is available data from the partial decay rates of
$D^{+} \to \bar{K}^{0} e^{+} \nu_{e}$ and $D^{0} \to K^{-} e^{+} \nu_{e}$ \cite{Besson:2009uv,Ge:2008aa}.
Given this experimental values and once the Wilson coefficients for THDM and Leptoquarks model has been computed,
we can constrain the values of the parameters in each model (THDM and Leptoquarks) allowed from the experimental data.
We will perform a $\chi^2$ analysis but using the BRs and furthermore, we will perform an analysis by using
ratios of the Branching ratios in order to eliminate the functional dependence on $V_{cs}$, which can be though as one of the
most important sources of error in the theoretical determination of the BR.
There are many possible combinations to define ratios of the BRs, for definitiveness we will define two
sets of ratios between the BRs as follow:
\begin{enumerate}
\item Set 1:
\begin{eqnarray}
R_1^{set_1}&=&\frac{\mathcal{B}^{th}_1(D^0\to K^- e^+ \nu_e)}{\mathcal{B}^{th}_2(D^0\to K^- \mu^+ \nu_\mu)},
R_2^{set_1}=\frac{\mathcal{B}^{th}_5(D_s^+\to \tau^+ \nu_\tau)}{\mathcal{B}^{th}_2(D^0\to K^- \mu^+ \nu_\mu)}\,,
R_3^{set_1}=\frac{\mathcal{B}^{th}_6(D_s^+\to \mu^+ \nu_\mu)}{\mathcal{B}^{th}_2(D^0\to K^- \mu^+ \nu_\mu)}\,,\nonumber\\
R_4^{set_1}&=&\frac{\mathcal{B}^{th}_3(D^+\to \bar K^0 e^+ \nu_e)}{\mathcal{B}^{th}_2(D^0\to K^- \mu^+ \nu_\mu)}\,,
R_5^{set_1}=\frac{\mathcal{B}^{th}_4(D^+\to \bar K^0 \mu^+ \nu_\mu)}{\mathcal{B}^{th}_2(D^0\to K^- \mu^+ \nu_\mu)}\,.\nonumber
\end{eqnarray}
\item Set 2:
\begin{eqnarray}
R_1^{set_2}&=&\frac{\mathcal{B}^{th}_1(D^0\to K^- e^+ \nu_e)}{\mathcal{B}^{th}_3(D^+\to \bar K^0 e^+ \nu_e)}\,,
R_2^{set_2}=\frac{\mathcal{B}^{th}_2(D^0\to K^- \mu^+ \nu_\mu)}{\mathcal{B}^{th}_3(D^+\to \bar K^0 e^+ \nu_e)}\,,
R_3^{set_2}=\frac{\mathcal{B}^{th}_5(D_s^+\to \tau^+ \nu_\tau)}{\mathcal{B}^{th}_3(D^+\to \bar K^0 e^+ \nu_e)}\,,\nonumber\\
R_4^{set_2}&=&\frac{\mathcal{B}^{th}_6(D_s^+\to \mu^+ \nu_\mu)}{\mathcal{B}^{th}_3(D^+\to \bar K^0 e^+ \nu_e)}\,,
R_5^{set_2}=\frac{\mathcal{B}^{th}_4(D^+\to \bar K^0 \mu^+ \nu_\mu)}{\mathcal{B}^{th}_5(D^+\to \bar K^0 e^+ \nu_e)}\,.\nonumber
\end{eqnarray}
\end{enumerate}
and in both cases we will also include ratios for the partial decay rates from CLEO, defined as follows ($i=1..10$ one per bin):
\begin{equation}
R_i^{D^0}=\frac{\Delta \Gamma_i(D^0\to K^- e^+ \nu_e)}{\Gamma(D^0\to K^- e^+ \nu_e)}\,,\quad
R_i^{D^+}=\frac{\Delta \Gamma_i(D^+\to \bar K^0 e^+ \nu_e)}{\Gamma(D^+\to \bar K^0 e^+ \nu_e)}\,.
\end{equation}
We perform a $\chi^2$ analysis for these 3 cases:
\begin{enumerate}
\item Using the total branching ratios
\begin{equation}
\chi^2=\sum_{i=1}^6\frac{(\mathcal{B}^{th}_i-\mathcal{B}^{exp}_i)^2}{\mathcal{\delta B}_i^2}+
\sum_{i=1}^{10}\frac{(\Delta \Gamma^{th_{D^0}}_i-\Delta \Gamma^{CLEO_{D^0}}_i)^2}{(\delta \Delta\Gamma_i^{D^0})^2}
+\sum_{i=1}^{10}\frac{(\Delta \Gamma^{th_{D^+}}_i-\Delta \Gamma^{CLEO_{D^+}}_i)^2}{(\delta \Delta\Gamma_i^{D^+})^2}\,.
\end{equation}
Here, $\mathcal{\delta B}_i$ is calculated adding in quadratures the experimental
and theoretical uncertainties shown in Table \ref{table1}.
\item And two different cases ($j=1,2$) for each selection of the ratios of the BRs and the ratios of the partial widths:
\begin{equation}
\chi^2_{set_j}=\sum_{i=1}^5\frac{R_i^{set_j}-R_i^{set_j^{exp}})^2}{(\mathcal{\delta} R^{set_j}_i)^2}+
\sum_{i=1}^{10}\frac{(R^{D^0}_i-R^{CLEO_{D^0}}_i)^2}{(\delta R_i^{D^0})^2}
+\sum_{i=1}^{10}\frac{(R^{D^+}_i-R^{CLEO_{D^+}}_i)^2}{(\delta R_i^{D^+})^2}\,.
\end{equation}
\end{enumerate}
\begin{figure}
\begin{center}
\includegraphics[width=0.76\textwidth]{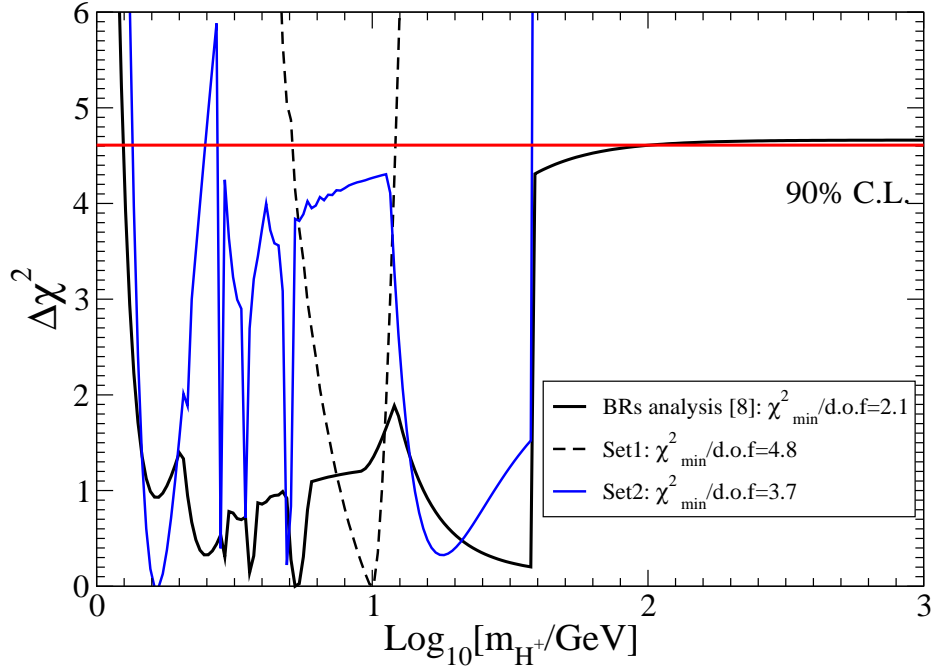}
\caption{$\Delta\chi^2$ projection on the mass of the charged Higgs within the Two Higgs Doublet Model type II with different
analysis.}\label{mhTHDMII}
\end{center}
\end{figure}
\subsection{Constraining Two Higgs Doublet Models with D meson decays}
Let us first show the results for the THDM type II. Other analysis have been done in the past,
but here we will show the bounds obtained by using D meson data only.
The projection over the mass of the charged Higgs is shown in Fig. \ref{mhTHDMII}.
Black solid line is the limit obtained by using the BRs \cite{Barranco:2013tba}, while blue and dashed line are obtained
using the ratios $R_i^{set_{1,2}}$ respectively. Although at 90\% C.L. there agreement of $set_1$ with the BRs analysis,
the analysis done with the ratios have a bad goodness of fit as expressed by $\chi^2_{min}/\mbox{d.o.f}$. Even
worst, the analysis done with $set_{1,2}$ seems to indicate that the mass of the charged Higgs should be always lower than
40 GeV. Although we would like to say that the THDM type II is discarded as direct searches performed by LEP have excluded
a charged Higgs with masses lower than 80 GeV. But this results depends on the election of ratios and the goodness of fit is
not reliable. Then, the limits we will report will be those coming from the analysis done with the BRs  only as it was done
in \cite{Barranco:2013tba}.
For the THDM type II, we found that a low mass for the charged Higgs is favored, at 90\% C.L. $6.3\mathrm{GeV} <m_{H^{+}}<63.1\mathrm{GeV}$, and at 95\% there is still agreement with the LEP constraints.
The reason of this region favored at 90\% is because, although there is an overall good agreement between theory and
experiment, the match is not perfect. See for instance the total branching
ratio labeled i=1 in Table 1 and shown in Fig. \ref{Brtheo}.
The theoretical BR and the world average are more than 1 sigma away. It can also be observed better in
the $q^2$ distribution shown in Fig. 2 of \cite{Barranco:2013tba} where most of the data points are above the theoretical decay rate.
Taking all observables together, and doing the analysis we have explicitly found and mentioned in the paper that the
significance of a light charged Higgs is at 90 \% C.L. and it comes from the small discrepancies between theory and experiment.
\begin{figure}
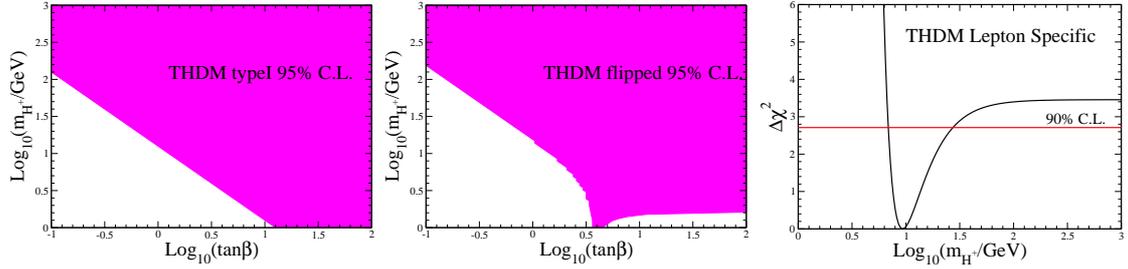

\begin{center}
\includegraphics[width=0.3\textwidth]{THDMI.eps}
\includegraphics[width=0.3\textwidth]{flipped.eps}
\includegraphics[width=0.3\textwidth]{LS.eps}
\caption{Bounds on different THDM types as expressed in Table \ref{table2}.}\label{THDMI}
\end{center}
\end{figure}

Finally, in Fig. \ref{THDMI} are shown the limits in the plane $(\tan\beta,m_{H^+})$ for the other THDM cases as illustrated in
Table \ref{table2}, namely the type I, the flipped and the lepton specific models (LS). Note that for the case
of the LS model, the values of $X,Y$ and $Z$ imply that the Wilson coefficients eq. \ref{WilsonTHDM} are independent of the
value of $\tan\beta$. This is the reason why it is shown only the projection over $m_{H^+}$ and the $chi^2$ analysis
show, as it is the case of the type II, a low mass for the Charged Higgs at 90\% C.L.

\subsection{Constraining Leptoquark scalar interactions with D meson decays}
In the case of the Leptoquark analysis, observe that the Lagrangian \ref{LeptoquarkLagrangian} includes Scalar, Vector and
Tensor NSIs which results in the Wilson coefficients eqs. \ref{WilsonLepto}. It is interesting to note that Scalar NSI are
proportional to the Tensorial components. Furthermore, the Vectorial component is real.
As we have already shown in Section \ref{Section2}, the decay rate $d\Gamma_{D\to K\ell\nu_\ell}/dE_K$ is not sensitive to the
tensor form factor as the tensor contribution is negligible when the scalar and vector interactions are taken into account
(see Fig. \ref{dgamma}), which are the dominant contributions. Hence the model dependent analysis is done varying 3 parameters at a time. An analysis including the tensorial contribution does not change our limits, hence they are independent of the election of $f_T$.
At 95\% C.L. the bounds for $C^{VLL}_{sc\ell\nu}$, expressed in GeV$^{-2}$ is given by
$0.04<|\kappa'^{L}_{i2}|^2/(m_{S_0}/300\mathrm{GeV})^2<0.11$, and the limits from $C^{SRR}_{sc\ell\nu}$ leads to the
following bounds in terms of the Leptoquark coupling constants and the mass os the scalar particles as:
\begin{eqnarray}
-0.17<\mathrm{Re}\left(\kappa^{R*}_{i2}\kappa^{L}_{i2}+\kappa'^{R*}_{i2}\kappa'^{L}_{i2}\right)/(m_{S}/300\mathrm{GeV})^2<0.01 , \nonumber \\
-0.09<\mathrm{Im}\left(\kappa^{R*}_{i2}\kappa^{L}_{i2}+\kappa'^{R*}_{i2}\kappa'^{L}_{i2}\right)/(m_{S}/300\mathrm{GeV})^2< 0.10\,.
\end{eqnarray}

\section{Conclusions}
We have shown that D meson decays may induce relevant constraints in some models as it is the case of the THDM and Leptquarks.
We have performed three different $\chi^2$ analysis. In two of them we have
used some selected ratios of the branching ratios (named $set_{1,2})$ and we have shown this two last types of analysis
have a couple of problems, even they look better as the ratios eliminate the dependence in the CKM matrix element $V_{cs}$,
which is a source of error. The analysis done with ratios of the Brs show that:
{\it i)} the goodness of the fit given by $\chi^2_{min}/\mbox{d.o.f}$ is worst when we use
the ratios instead of the decay rates, and {\it ii)} given the arbitrary of selecting different ratios, the bounds change.
Hence, we believe the analysis done with the BRs only, as those are measured quantities implies reliable bounds. For further
constraints in a model independent analysis or for other models beyond SM, we refer to the analysis done in \cite{Barranco:2013tba}.
The bounds presenter here for the THDM and Leptoquark models are complementary to those presented in  \cite{Eberhardt:2013uba}.

\section*{Acknowledgments}
This work has been supported by CONACyT SNI-Mexico. The authors are also grateful to Conacyt (M\'exico) (CB-156618), DAIP project (Guanajuato University) and PIFI (SEP, M\'exico) for financial support.

\end{document}